\documentclass{article}

\usepackage{arxiv}

\usepackage[utf8]{inputenc} 

\usepackage{amsthm,amsmath}

\usepackage{booktabs} 
\usepackage[autostyle]{csquotes}
\usepackage{xspace}
\usepackage{tabularx}
\usepackage{graphicx}
\usepackage{xcolor}
\usepackage{balance}
\usepackage{multirow}
\usepackage{hyperref}
\usepackage{natbib}
\usepackage[labelfont=bf]{caption}
\setcitestyle{square,sort,comma,numbers}

\newcommand{\revision}[1]{\textcolor{black}{#1}}

\newcommand{\uk}{{\sc UK}\xspace}

\newcommand*\justify{%
	\fontdimen2\font=0.4em
	\fontdimen3\font=0.2em
	\fontdimen4\font=0.1em
	\fontdimen7\font=0.1em
	\hyphenchar\font=`\-
}
\newcommand{\soft}[1]{\texttt{\justify #1}}

\newcommand*{\SuperScriptSameStyle}[1]{%
  \ensuremath{%
    \mathchoice
      {{}^{\displaystyle #1}}%
      {{}^{\textstyle #1}}%
      {{}^{\scriptstyle #1}}%
      {{}^{\scriptscriptstyle #1}}%
  }%
}

\newcommand*{\oneS}{\SuperScriptSameStyle{*}}
\newcommand*{\twoS}{\SuperScriptSameStyle{**}}

\title{Responsible team players wanted: an analysis of soft skill requirements in job advertisements}

\author{
Federica Calanca\\
  Sapienza University of Rome\\
  Italy \\
  \texttt{fede.calanca@gmail.com} \\
   \And
  Luiza Sayfullina \\
  Aalto University\\
  Finland\\
  \texttt{sayfullina.luiza@gmail.com} \\
  \And
  Lara Minkus\\
  University of Bremen\\
  Germany\\
  \texttt{lminkus@bigsss.uni-bremen.de } \\
  \And
  Claudia Wagner \\
  GESIS \& University Koblenz \\
  Germany \\
  \texttt{claudia.wagner@gesis.org} \\
  \And
  Eric Malmi \\
  Aalto University\\
  Finland\\
  \texttt{eric.malmi@gmail.com} \\
}
\date{}

\begin{document}
\maketitle

\begin{abstract}
During the past decades the importance of soft skills for labour market 
outcomes has grown substantially. This carries implications for labour market 
inequality, since previous research shows that soft skills are not valued equally across race and gender. 
This work explores the role of soft skills in job advertisements by drawing on methods from 
computational science as well as on theoretical and empirical insights from 
economics, sociology and psychology. We present a semi-automatic approach 
based on crowdsourcing and text mining for extracting a list of soft skills.
We find that soft skills are a crucial component of job ads, especially of low-paid jobs and jobs in female-dominated professions.
Our work shows that soft skills can serve as partial predictors of the gender composition in job categories and that not all soft skills receive equal wage returns at the labour market. Especially ``female'' 
skills are \revision{frequently} associated with wage penalties. 
Our results expand the growing literature on \revision{the association of soft skills on wage inequality} and highlight \revision{their} importance for occupational gender segregation at labour markets.
\end{abstract}

\keywords{
soft skills\and
job advertisement\and
text mining\and
computational social science\and
crowdsourcing\and
gender inequality\and
labour markets
}

\section{Introduction}
\label{sec:intro}

When it comes to jobs and careers, technical abilities and professional qualifications are important factors both from the perspective of an employer and of a new employee. However, as pointed out by recent studies\revision{~\cite{mcdonald, bacolod2010two, Alabdulkareemetal2018}}, more and more attention is focused on soft skills, i.e. qualities that do not depend on the acquired knowledge and that are harder to quantify due to being related to one's emotional intelligence and personality traits. At the same time, they are extremely important \revision{because they} facilitate human connections~\cite{monster_report}. The Oxford dictionary, for instance, defines soft skills as ``personal attributes that enable someone to interact effectively and harmoniously with other people''\footnote{\url{https://en.oxforddictionaries.com/definition/soft_skills}}.
During the period of 1980 and 2012, jobs with high social skills requirements grew by around 10\% as a share of the US labour force~\cite{future2030}. \revision{The increasing} importance of soft skills at labor markets stems from the growth of the service sector, where interpersonal services are sold, as well as from the introduction of lean-manufacturing, where an integrated skill set, comprised of both hard and soft skills, has gained importance~\cite{grugulis2009whose,shibata2001productivity}. 

\revision{Observational studies have also shown} that social features \revision{potentially} related \revision{to} soft skills (e.g. the variety of friendship connections and position diversity within a community) are positively correlated with economic outputs\revision{~\cite{xie2017, wachs2017}}. 

The growing importance of soft skills also carries implications for gender inequality in labour markets. Research has shown that certain societal groups are perceived as lacking important soft skills, i.e. evidence was found that black men are characterized as being less motivated than their white counterparts~\cite{moss1996}. Additionally, not all types of soft skills are valued equally, e.g. based on gender stereotypes and beliefs about women's inferior status in the workplace, skills that are perceived as ``female'' are found to be associated with wage penalties~\cite{ridgeway1997,Ceci22022011,lester2010}. \revision{On the other hand, recent scholarly debates engage in the discussion of a possible female advantage associated with the rising importance of people skills in contemporary labor markets~\cite{bacolod2010two, black2010explaining, balcar2014soft, borghans2014}}.

Despite the growing importance of soft skills and their \revision{potential contributions} to inequalities in labour markets, to date, we know surprisingly little about the role of ``\revision{gendered soft skills'' -- i.e., soft skills that are stereotypically associated with one gender --} in the job market\revision{~\cite{balcar2014soft, levanon2016persistence}}. 
\revision{Most prior scientific articles referring to skills and labor market outcomes construct indices of soft skills in which male and female connoted skills get added up, rather than making a distinction between them (see, for instance,~\cite{bacolod2010two, black2010explaining}). This approach is useful, because the overall increasing importance of soft skills in contemporary labor markets~\cite{borghans2014} can be measured in an easy-to-grasp, single-index way. However, this coarse-grained measure can mask important differences in labor market outcomes with regard to gendered soft skills. \\We go beyond this relatively crude measure by introducing a semi-automatic approach for constructing an extensive list of soft skills from job advertisements, which we can use for soft skills detection. Combining this data on soft skills with what prior research has identified as commonly shared gender stereotypes (see, for instance, ~\cite{bem1974, gaucher2011evidence, rudman2001}) and official statistics about the proportion of women in various professional fields, allows us to differentiate soft skills depending on their gender connotation. Thus we are able to establish new insights on the association of soft skills related to gender stereotypes and wages.} 

Additionally, we present evidence on the impact of soft skills on sex segregation \revision{in} labor markets. 
Although the existing literature on supply-side mechanisms of occupational sorting, i.e. women making career choices based on potentially biased self-assessed beliefs about interests and capacities, is growing~\cite{busch2014supply,correll2001}, the demand-side process, meaning the allocation of men and women into sex-typed occupations by employers, remains \revision{relatively understudied}~\cite{levanon2016persistence}. 
\revision{There is only a limited number of studies examining the influence of gendered wording on occupational choices. These studies use small-scale experiments and thus cover only a limited range of soft skills associated with gender stereotypes~\cite{askehave2014gendered, gaucher2011evidence, bem1973does, taris1998gender, born2010impact}}. Utilizing our newly extracted dataset based on real job advertisements, we are able to examine the impact of soft skills in general and gendered soft skills in specific on occupational segregation. 

\revision{Based on our unique dataset on soft skills in job ads, we find evidence that female connoted soft skills are associated with wage penalties, while soft skills perceived as being stereotypically male are linked to wage premiums. Our results show further that women are more likely to be found in occupations that are advertised using soft skills associated with female stereotypes and vice versa for men.} 

\revision{This article} is structured as follows: \revision{i}n Section~\ref{sec:mining}, we present our methodology for extracting soft skill mentions from a large corpus of job advertisements. In Section~\ref{sec:salary}, we scrutinize wage premiums and penalties associated with soft skills frequently mentioned in job ads based on a matching study. Next, the role of soft skills in reproducing gender segregation, \revision{i.e. the unequal distribution of men and women across occupations}, is examined in Section~\ref{sec:gender}. Finally, we present conclusions in Section~\ref{sec:conclusions} with a summary of our findings, their implications, limitations, and suggestions for future work.

\section{Methods and Data }
\label{sec:mining}

In this section, we describe the datasets used in this work and our semi-automatic soft skill mining approach. \revision{Following this approach we first create clusters of soft skills, grouping similar soft skills together, and then detect soft skills in job ads by searching for the soft skill strings in job descriptions.}

\subsection{Data}
\revision{Our analysis} is based on a dataset \revision{containing} 245,000 job advertisements (ads) from the United Kingdom (UK).\footnote{The dataset from UK is available at: \\ \url{https://www.kaggle.com/c/job-salary-prediction}} This data is provided by the Adzuna job search engine, which collects job ads from hundreds of different websites. Each job ad entry contains the \textit{title}, \textit{full description}, \textit{job category}, and \textit{salary} of the job, among five other types of fields.\footnote{\revision{Additional variables of the dataset encompass: location, type of contract (full- vs. part-time), length of contract (contract-based vs. permanent), the company name, and the source of the job ad.} }

\revision{Adzuna has classified the ads into 29 job categories}
, based on the source of the ad and the job's description. \revision{Table~\ref{tab:distinctiveness} illustrates the most distinctive soft skills for five selected job categories}. Desired soft skills differ considerably depending on the \revision{job category}. For instance, the three most distinctive skills for \textit{Teaching} are \soft{enthusiastic, dedicated, professional}, whereas for \textit{Accounting \& Finance} they are \soft{accurate, responsible, analytical abilities}. The soft skill detection algorithm is described in Section~\ref{ssdetection}. 

\begin{table*}[ht]
\centering
\caption{
The most distinctive soft skills for five job categories. Distinctiveness ($\Delta$ \%) is defined as the absolute difference between the percentage of job ads that contain the skill within the given category (\%) and the percentage of job ads that contain the skill within all categories.}
\label{tab:distinctiveness}
\resizebox{\textwidth}{!}{
\begin{tabular}{lcc|lcc|lcc|lcc|lcc}
\hline
\textbf{Social work} & $\Delta$ \% & \% & \textbf{Accounting \& Finance} & $\Delta$ \% & \% & \textbf{IT} & $\Delta$ \% & \% & \textbf{Teaching} & $\Delta$ \% & \% & \textbf{Creative \& Design} & $\Delta$ \% & \%  \\ \hline
team player & +7.3 & 22.7 & accurate & +7.5 & 14.1 & problem solving & +4.6 & 8.9 & enthusiastic & +12.1 & 20.2 & creative & +24.8 & 30.3 \\
ability to work with children & +6.6 & 7.0 & responsible & +6.0 & 34.7 & communication skills & +3.5 & 27.8 & creative & +5.9 & 11.5 & innovative & +5.3 & 11.2 \\
positive & +4.3 & 9.8 & communication skills & +4.6 & 28.9 & innovative & +3.1 & 8.9 & positive & +5.9 & 11.4 & attention to detail & +4.9 & 9.8 \\
flexible & +1.5 & 13.4 & analytical skills & +3.2 & 5.9 & team player & +2.4 & 17.8 & leadership & +5.0 & 11.4 & management skills & +4.4 & 14.2 \\
leadership & +1.5 & 7.9 & attention to detail & +2.9 & 7.8 & analytical skills & +2.1 & 4.8 & confident & +4.2 & 11.1 & responsible & +3.6 & 32.4 \\
patience & +0.8 & 0.9 & ability to work within deadlines & +2.1 & 4.7 & management skills & +1.8 & 11.6 & hard working & +3.4 & 6.3 & confident & +3.0 & 9.8 \\
people skills & +0.8 & 2.3 & interpersonal skills & +1.5 & 6.0 & creative & +1.5 & 7.0 & innovative & +3.1 & 8.9 & presentation skills & +1.6 & 3.4 \\
\hline
\end{tabular}
}
\end{table*}

All experiments in this paper are conducted using the \uk dataset, except for a crowd-sourcing experiment needed for collecting an initial list of soft skills, which is described in the next Section~(\ref{sec:crowd}). For this crowd-sourcing experiment, a dataset \revision{posted by the Armenian human resource portal CareerCenter consisting of 19,000 online job postings in a period from 2004--2015 is more appropriate, because job requirements are listed in a separate field.  Thus the workers do not need to read through the full ad, allowing us to annotate more ads and to collect a longer list of soft skills.}\footnote{The Armenian dataset is available at: \\ \url{https://www.kaggle.com/madhab/jobposts} \\
\revision{Using  a  different  dataset  carries  the  risk  that  some skills might only appear in the UK dataset}. However, this  most  likely  only  applies  to  very  infrequent  soft skills  and  \revision{thus would} have  little  effect  on  the  down-stream analyses.}

\subsection{Soft Skill Mining}
\label{sec:mining2}
\begin{figure*}[ht!]
    \centering
    \includegraphics[width=0.95\textwidth]{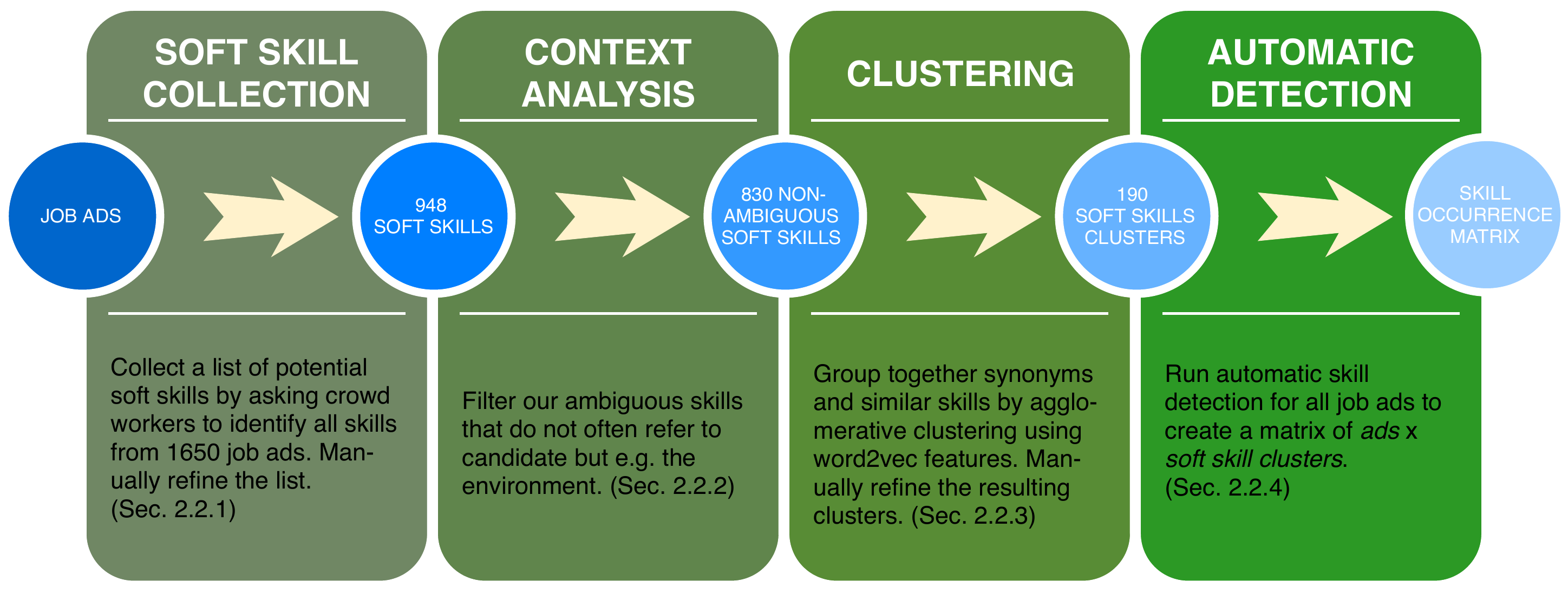}
    \caption{The steps of our data-extraction process.
    We collect a list of soft skill clusters using crowd sourcing and then find occurrences of these clusters in a corpus of job ads.}
    \label{fig:method}
\end{figure*}

Our semi-automatic soft skill mining approach consists of the following steps: first, crowdworkers generate an initial set of potential soft skills, second, skills that seldom refer to candidates are removed, third, soft skills with a similar meaning are clustered into groups of skills, and fourth, soft skills are detected in new ads. These steps are summarized in Figure~\ref{fig:method} and explained in more detail in the following sections.

The resulting soft skills and their clusters are available at \url{https://drive.google.com/drive/folders/1N1XkmgJ8awB9SgQjcdsYqMK7oQJAqtKo?usp=sharing}.

\subsubsection{Crowdsourcing a List of Soft Skills}
\label{sec:crowd}

The collection of soft skills was done through Figure Eight (formerly known as CrowdFlower)\footnote{\url{https://www.figure-eight.com/}}, a crowdsourcing platform that allowed us to speed up our data collection process by submitting annotation tasks to online crowdworkers.

First, each worker was given the following definition of soft skills:
\blockquote{
In a nutshell soft skills can be identified as qualities that do not depend on acquired knowledge; they complement hard skills (also known as technical skills).
According to Wikipedia soft skills ``are a combination of interpersonal people skills, social skills, communication skills, character traits, attitudes, [$\ldots$] social intelligence and emotional intelligence quotients'' 
}
This was followed by a list of soft skill examples and instructions for completing the tasks.
In particular, the workers were instructed to read the presented text, consisting of the ``job description'' and ``required qualifications'' fields, select whether the text contained any soft skills, and, if that was the case, they were instructed to copy and paste the smallest relevant part of text denoting each skill to an answer field. Additionally, the workers were instructed to remove unnecessary adjectives and complements, but not to alter the text in any other way. For instance, \soft{Excellent communication skills with customers and partners} had to be reported as \soft{communication skills}.

Before the actual annotation phase, the workers were supposed to pass a training phase and answer a set of test questions, for which we had provided the correct answers: they had to obtain an accuracy level of at least 60\% to proceed further. These test questions also showed up randomly during the actual annotation phase to ensure that the minimum accuracy level of 60\% was maintained.

\begin{figure}[t!]
    \centering
    \includegraphics[width=0.7\textwidth]{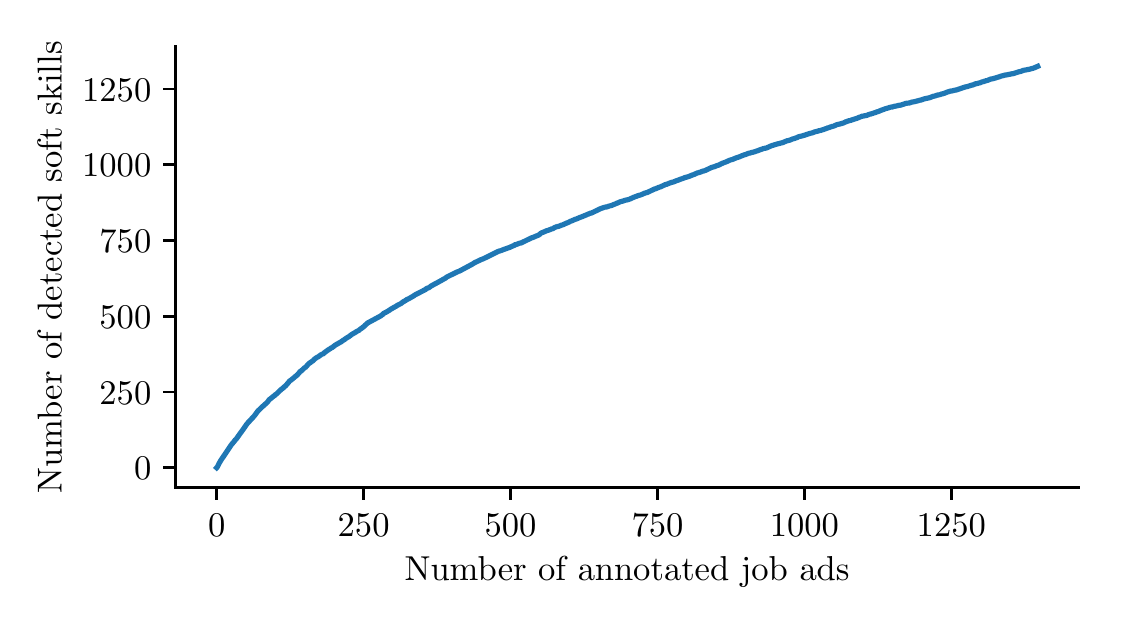}
    \caption{The cumulative number of discovered soft skills as function of annotated job ads.
    The rate of discovered soft skills slows down towards the end of our data collection. At the end, the newly discovered skills are mostly typos and other phrases unrelated to soft skills. The final, manually refined list consists of 948 unique soft skills.}
    \label{fig:skill_counts}
\end{figure}

In total, we annotated 1,650 job ads by at least 3 different workers.
The annotation effort was conducted in two batches. \revision{After both batches we computed the number of distinct soft skills as a function of the number of annotated ads, plotted in Figure~\ref{fig:skill_counts}}. The results show that the rate at which new soft skills are discovered \revision{slows} down, although new skills were still found at the end of the data collection. However, when examining the skills found last, most of them turned out to be typos and \revision{other phrases unrelated to soft skills (these include ``ability to work as a part of PSD team'', which is a hard skill since PSD stands for \textit{personal security detail}, and ``unquestioned behaviour'', which is highly ambiguous). 
Therefore, }we decided to stop the annotation task after the second batch.

To remove the typos as well as recurrent superfluous adjectives\revision{\footnote{\revision{The list of superfluous adjectives includes: \normalsize\soft{excellent}, \soft{highly}, \soft{very good}, \soft{good}, \soft{strong}, and \soft{high}.}}}, results were cleaned using a script. The script removed additionally extra whitespace and punctuation, and it corrected simple typos and misspellings by comparing the detected skill tokens to a whitelist of valid skill tokens. Thereafter, we manually reviewed the skills to remove all non-soft skills and to prune out tokens not relevant to the skill.

The final manually curated collection included 948 unique soft skills.

\subsubsection{Removing Ambiguous Soft Skills}
\label{sec:ambig}

The focus of this work is to analyze soft skill requirements for job applicants. However, often soft skill phrases in job ads do not refer to the required applicant characteristics, but they may also describe the working environment or something else. For instance, \soft{independent} could be used to describe an ``\textit{independent} business'' or a \textit{home care assistant} might be required to ``help people to remain \textit{independent} in their own homes.'' Therefore, it is crucial to be able to detect soft skills that refer to the candidate rather than something else.

To tackle this problem, we created another crowdsourcing task, instructing crowdworkers to annotate soft skill phrases in the context they appear, i.e. the job ads. We noticed that skills consisting of multiple tokens usually unambiguously refer to the candidate
and therefore we only annotated the skills consisting of at most three words, that is, 582 out of the 948 skills found in the previous steps. 

More specifically, for each one of these skills, we extracted 10 randomly sampled text snippets where the skill occurs, including 25 words before and after the skill. 
Then we asked crowdworkers to classify each snippet to one of the following three categories: \textit{Candidate}, \textit{Company/Company environment}, or \textit{Other}. At least three answers were recorded for each text snippet.

Based on the annotations, we computed the following confidence score\footnote{\url{http://success.crowdflower.com/hc/en-us/articles/201855939-How-to-Calculate-a-Confidence-Score}} for each soft skill
\begin{equation*}
\mathrm{Conf}(s) = \frac{\sum_{w \in W_c(s)}T(w) }{\sum_{w \in W(s)}T(w)} \,,
\end{equation*}
where $W_c(s)$ denotes the workers who classified an occurrence of skill $s$ to refer to a candidate, $W(s)$ denotes the workers who assessed an occurrence of skill $s$, and $T(w)$ is the trust of a worker $w$. \revision{Trust} is calculated by the crowdsourcing platform as the contributor's accuracy level in the current job, determined by his/her accuracy during the training phase -- as explained in Section~\ref{sec:crowd}.
Thus, the confidence score measures the proportion of votes for the \textit{Candidate} category weighted by the trusts' of the workers who gave the votes.

We included the skills with a confidence value of at least 0.7 into the final list of soft skills. This value allowed us to retain 81.3\% of the annotated skills (8.3\% of trigram, 10.3\% of bigram and 40.1\% of single-word skills were discarded) while still having a relatively high confidence that the retained soft skill phrases actually refer to the candidate.

\subsubsection{Soft Skill Clustering}
Many of the soft skills collected by the crowdworkers are synonyms or near-synonyms. The different versions of a skill result, e.g., from diverse ways of expressing the concept (\soft{team-worker}, \soft{ability to work in a team}), or from slightly different spellings (\soft{able to work in team}).
To unify the different variants, the collected soft skills were clustered by first employing an algorithmic approach and then refining the clusters manually.
After experimenting with a small subset of soft skills, different algorithms and parameter settings, we decided upon the following procedure.

Each soft skill was first represented in the vector space by averaging the word2vec~\cite{mikolov2013} embeddings of its tokens, excluding stopwords. 
We used 300-dimensional embeddings pre-trained on the GoogleNews dataset.\footnote{Official archive available at: \\ 
	\url{https://drive.google.com/file/d/0B7XkCwpI5KDYNlNUTTlSS21pQmM}}

Then, we employed agglomerative clustering algorithm to cluster the embedding vectors using the \textit{average linkage} cosine distance measure. The clusters were finally reviewed and manually improved by split and merge operations and by reassigning some of the skills to more appropriate clusters, obtaining a final list of 190 clusters.\footnote{\revision{Clusters are available
at: \url{https://docs.google.com/spreadsheets/d/1ke-xjXfwYHG-KPrEnJ90v41A7VzO82P7cx3Ez4lq9fQ/edit?usp=sharing}} }

\subsubsection{Soft Skill Detection}\label{ssdetection}
In the final phase, our goal was to detect skill clusters in each job ad.

First, we preprocessed the job descriptions and the list of soft skills by lowercasing and removing stop words.\footnote{We used the list of English stop words from the NLTK package (\url{http://www.nltk.org}).}
We also removed the competence terms (\soft{able, skills, etc.}) from most soft skills, if they were perceived as not being fundamental for skill identification, to avoid false negatives (e.g. \soft{capable of handling multiple tasks} should match with \soft{abilities in handling multiple tasks}). Still, for some skills, we kept the competence terms if they would have become too ambiguous, resulting in false positive detection (e.g. 
\soft{communication skills} without the word \soft{skills} would match with communication technologies).

Thereafter, we searched for each soft skill $s$ in each job description. If $s$ consisted of multiple tokens, we allowed for at most two extra words to occur before each token in addition to stop-words, that were allowed to be removed from certain skills without making them ambiguous. \revision{We also experimented with more liberal ways of matching skills, ignoring the word order of the skill tokens or lemmatizing the tokens, but these were found to decrease the precision of the detected skills significantly.} 

Soft skills were detected in 78\% of the ads, with 45.5\% mentioning at least 3 soft skills, attesting to the importance of soft skills in the labour market.

\subsubsection{\revision{Related Work on Soft Skill Mining}}

\revision{The curation of \textit{hard} skills has been addressed by LinkedIn~\cite{bastian2014linkedin}, whereas Kivim\"{a}ki et~al.~\cite{kivimaki2013graph} proposed a system for automatic detection of new skills in free written text using a spread-activation algorithm. 
Recently, Haranko et al.~\cite{haranko2018professional} suggested a novel approach for collecting data on skills and gender imbalances through LinkedIn's advertising platform.
Automatic classification of soft skills referring to a candidate vs. something else (e.g. the work environment), has been studied by Sayfullina et al.~\cite{sayfullina2018learning}, using the crowdsourced data collected in this work as described in Section~\ref{sec:ambig}.}

\section{Salary and Soft Skills}
\label{sec:salary}

One of our main research questions is how the presence of certain soft skills may affect wages. 

\begin{figure}[t!]
    \centering
    \includegraphics[width=0.6\textwidth]{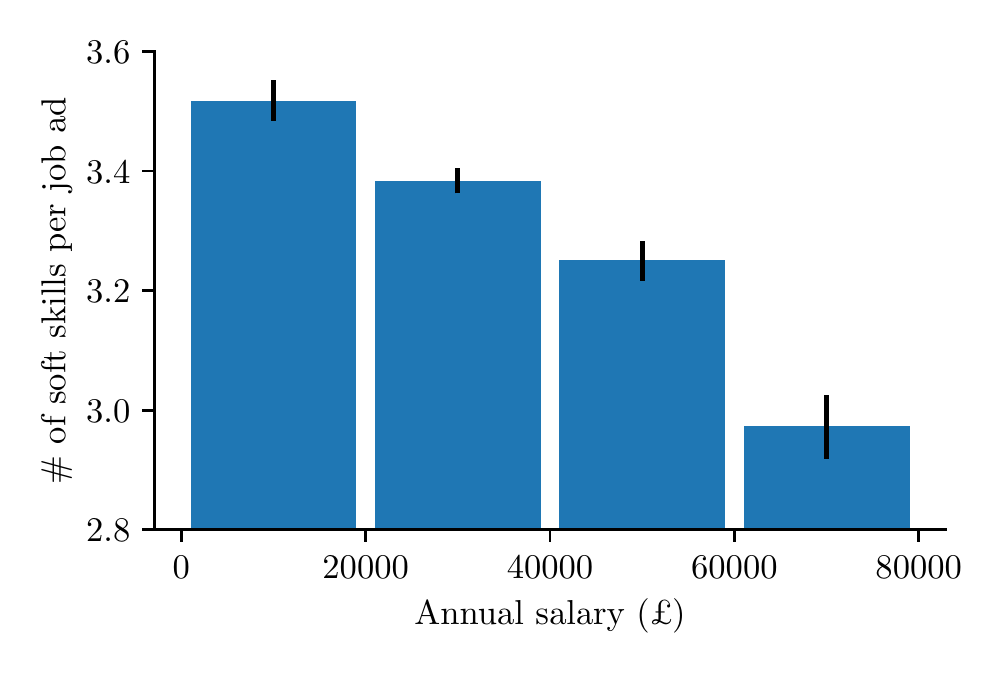}
    \caption{Low-paid job ads contain, on average, more soft skills than high-paid job ads. Bars show the average number of soft skills for the ads in four different salary groups, and the error bars indicate the 95\% confidence intervals obtained via bootstrap re-sampling with replacement.
    }
    \label{fig:skill_counts_by_salary}
\end{figure}

\revision{Analyzing annual salaries of job ads, we found that low-paid job ads contain, on average, more soft skills than high-paid job ads. This is illustrated in Figure~\ref{fig:skill_counts_by_salary} which shows the average number of soft skill mentions per job ad in four different salary groups. The ads with a salary ($s$) of $s \leq \pounds 20,000$ have 3.52 soft skills on average, whereas ads with a salary of $\pounds 60,000 < s \leq \pounds 80,000$ have only 2.97 soft skills on average. All paired differences between the salary groups are statistically significant (\revision{$p < 0.001$; two-tailed $t$-tests with unequal variances).}} 
   
While the higher prevalence of soft skills in low paid jobs is interesting by itself, it does not reveal which soft skills tend to be associated with wage premiums and which ones with wage penalties.
To address this question we conduct a matching study.

\subsection{Matching Study}
\newcommand{\R}{\ensuremath{r}\xspace}
\newcommand{\boldR}{\ensuremath{\mathbf{r}}\xspace}
\newcommand{\AllCat}{\mathcal{A}}
\newcommand{\Cat}[1]{\mathcal{A}\left(#1\right)}
\newcommand{\T}{\ensuremath{T}\xspace} 
\newcommand{\Mw}{\ensuremath{M}\xspace}
\newcommand{\Mwo}{\ensuremath{\bar{M}}\xspace}
\newcommand{\Cw}{\ensuremath{C}\xspace}
\newcommand{\Cwo}{\ensuremath{\bar{C}}\xspace}

In order to study the link between a job ad's soft skill requirements and \revision{their} respective salary,\footnote{The job ads do not mention the exact annual salary but only a range, so we use the median of the range as the job salary.} we conduct a matching study~\cite{rosenbaum1983central}. 
The benefit of matching is that, in pairing a treated job ad (i.e. an ad with a given job title and job category that contain a specific skill) with its counterfactual (i.e. an ad with the same title and category but without the specific skill), we can control for a range of unobserved job category characteristics~\cite{angrist2008mostly}. \revision{These characteristics include, for instance, work experience, since job titles often include qualifiers, such as \textit{head}, \textit{senior}, \textit{junior}, or \textit{intern}.}

The specific matching strategy applied in this article is as follows:
first, we group ads having the same job category $c$ and job title $t$, ignoring stop words and the word order of the title. We picked all titles occurring at least twice, resulting in 34,071 distinct titles and 158,658 ads. 
Given a soft skill $s$, a normalized salary reward is defined as
\begin{equation} \label{eq:r_ct}
   \R_{s,c,t} = \frac{\Mw_{s,c,t} - \Mwo_{s,c,t}}{\Mwo_{s,c,t}} \times 100\% \,,
\end{equation}
where $\Mw_{s,c,t}$ and $\Mwo_{s,c,t}$ are the average salaries of job ads belonging to job category $c$, having job title $t$, and containing or not containing skill $s$, respectively.

\revision{
For example, in our dataset there are 210 ``Java Developer'' job ads in the IT Jobs category out of which 28 contain the soft skill \soft{communication skills}. The average salary of these 28 positions is \pounds46,536 per year, whereas the average salary for the other 182 positions is \pounds43,170 per year. This means that the salary reward for \soft{communication skills} in Java Developer / IT Jobs category is
\begin{align*} 
   &\R_{\textmd{communication skills, IT Jobs, Java Developer}} \\
   &\qquad= \frac{\textmd{\pounds}46,536 - \textmd{\pounds}43,170}{\textmd{\pounds}43,170} \times 100\% \\
   &\qquad= 7.8\%\,,
\end{align*}
suggesting that Java developer positions that require communication skills usually pay $7.8\%$ more than other Java developer positions.
}

Given the individual salary rewards, the overall salary reward $\R_s$ of soft skill $s$ is obtained by averaging the rewards over all possible job titles and categories
\begin{equation} \label{eq:r}
   \R_s = \frac{\sum_{c} \sum_{t} r_{s,c,t} \min\left(\Cw_{s,c,t}, \Cwo_{s,c,t}\right)}{\sum_{c} \sum_{t} \min\left(\Cw_{s,c,t}, \Cwo_{s,c,t}\right)} \,,
\end{equation}
where $\Cw_{s,c,t}$ and $\Cwo_{s,c,t}$ are the number of job ads belonging to job category $c$, having job title $t$, and containing or not containing skill $s$, respectively. 
Individual rewards are weighted by the number of ads to avoid letting infrequent job titles have disproportionately large effect on the overall reward. In most cases, $\min\left(\Cw_{s,c,t}, \Cwo_{s,c,t}\right) = \Cw_{s,c,t}$ since typically less than half of the ads from any category contain a given soft skill. Thus, the individual rewards are typically weighted by the number of ads containing the skill.

A positive reward $\R_s$ indicates that job ads that mention skill $s$ have on average a higher salary than other job ads from the same job category and the same job title that do not mention $s$.

\revision{To compute the statistical significance of an observed reward value, $\R^\textmd{obs}$, we conduct a permutation test as follows: each job ad consists of ($i$)~a set of soft skills mentioned in the job description, ($ii$)~job category and title, and ($iii$)~salary. We shuffle the soft skill sets ($i$) between the ads and keep everything else ($ii$ and $iii$) fixed.
This shuffling is repeated 1,000 times and after each shuffle, we compute a new reward $\R^\textmd{rand}$. The $p$-value for the null hypothesis that $|\R^\textmd{obs}| \leq |\R^\textmd{rand}|$ is given simply by the fraction of $|\R^\textmd{rand}|$ values that are greater than or equal to $|\R^\textmd{obs}|$. If the fraction is below or equal to a threshold of $\alpha=0.05$, we conclude that $\R^\textmd{obs}$ is statistically significant and mark the reward with a '\oneS'. A reward with $p \leq 0.01$ is marked by '\twoS'.} 

\subsection{Results}

The soft skills that are associated with the highest wage premiums or penalties are shown in Table~\ref{top_10_rewarding_skills}.
Most of the soft skills associated with wage premiums can also be considered a requirement for higher occupational positions. Soft skills such as \soft{delegation skills}, \soft{team building skills} and \soft{leadership} imply that a certain kind of supervision and authority toward others is required~\revision{\cite{goldthorpe2007sociology}}. In contrast, \soft{listening skills, willingness to learn}, as well as being \soft{punctual}, describe skills that entail a certain degree of subordination.

Our empirical observation that soft skills associated with wage premiums are also closely tied to leadership positions is in accordance with sociological occupational class theories. Previous research on occupational classes has identified the magnitude of a job's authority as one of the key determinants in assessing the job's position in the occupational class system\revision{~\cite{hertel2016,wright1997}}. Jobs that entail a high  degree of authority also occupy a strategic position in the labour market: by monitoring their subordinates, employees in leadership positions are ensuring that a firm produces surplus. Given this powerful position, high degrees of authority entail a significant degree of bargaining power and thereby the possibility to demand higher than average wages~\cite{wright1997}. \revision{Empirical research indeed supports this notion and shows that \soft{leadership} skills are associated with wage premiums~\cite{kuhn2005, weinberger2014}}.

\begin{table}[ht]
\centering
\caption{Skills with the highest and the lowest overall salary rewards (\boldR from Eq.~\ref{eq:r}).
The \textit{Count} column shows the denominator from Eq.~\ref{eq:r}, which can roughly be interpreted as the sample size. Only skill clusters with \textit{Count} $\geq 50$ are shown.} 

\label{top_10_rewarding_skills}
\begin{tabular}{l l r}
\toprule
\textbf{Skill Cluster}  & \boldR & \textbf{Count} \\
\midrule
maturity & 11.9\twoS& 112\\
delegation skills & 10.2\twoS& 53\\
team building skills & 9.8\oneS& 50\\
strategic planning & 9.1\twoS& 608\\
ability to work  in a fastpaced environment & 8.0\oneS& 51\\
leadership & 7.4\twoS& 4743\\
constructive feedback & 6.9\oneS& 74\\
proposal writing & 6.2\oneS& 84\\
ability to improve skills & 6.0\twoS& 108\\
discretion & 5.7& 309\\
results driven & 4.9\twoS& 541\\
presentation skills & 4.5\twoS& 1464\\

\midrule 
telephone skills & -7.3\twoS& 227\\
polite & -5.9\twoS& 339\\
dynamic person & -5.2& 70\\
dedication & -4.6\twoS& 467\\
friendly personality & -4.6& 97\\
listening skills & -4.3\twoS& 355\\
punctual & -4.1\oneS& 248\\
ability to identify problems & -3.1& 132\\
calm & -2.8\oneS& 787\\
professional manner & -2.6\twoS& 2303\\
willingness to learn & -2.2\twoS& 1652\\
time management & -1.8& 2149\\
 
\bottomrule 
\addlinespace[1ex]
\multicolumn{3}{l}{\twoS$p<0.01$, \oneS$p<0.05$}
\end{tabular}
\end{table}

Additional supporting evidence for this particular reading of the results comes from psychology. We find that character traits associated with wage premiums, for instance \soft{delegation skills}, \soft{team building skills}, and \soft{strategic planning} are closely connected to skills psychological research has identified as leadership characteristics, i.e. \texttt{management of personnel, visioning}, as well as general \texttt{strategic skills}~\cite{mumford2007}. 

What is striking, is that many of the aforementioned skills in Table~\ref{top_10_rewarding_skills} also correspond to gender stereotypes. Gender stereotypes are generalizations about commonly shared perceptions of female and male attributes. Previous research has shown that while women are described as embodying ``communal behavior'', such as \soft{kindness, loyalty}, and \soft{warmness}, men are characterized by ``agentic traits'', such as \soft{competitiveness} and \soft{aggressiveness}~\cite{rudman2001}, and as possessing \soft{leadership abilities}~\cite{bem1974}. Common ``agentic'' traits, such as \soft{competitive} and \soft{aggressive}, have been filtered out as ambiguous (see Section~\ref{sec:ambig}), since they typically do not describe the desired characteristics of the job applicant. However, we \revision{still} find several leadership traits to come about with higher wages in Table~\ref{top_10_rewarding_skills}. Moreover, ``communal behavior'' seems to come about with wage penalties in Table~\ref{top_10_rewarding_skills} across the board (for instance: \soft{polite, dedication, friendly personality}, and being \soft{calm}). 

Thus, Table~\ref{top_10_rewarding_skills} provides evidence that male gender stereotypes are connected to wage premiums, whereas female gender stereotypes are connected to wage penalties in the labor market. To scrutinize this issue further, \revision{in the following} section we examine the association between gender stereotypes and wages in more detail.

\section{Gender and Soft Skills}
\label{sec:gender}
\revision{In this section we scrutinize to what extent soft skills are associated with occupational sex segregation. Thereafter, we explore a possible relationship between wages and gendered soft skills.}

\subsection{Industry Gender Composition Prediction}\label{industry_gender_decomp}
In what follows, we test whether soft skills can predict the gender composition of a job category. The proportion of women for each job category was approximated by mapping the job categories in our data to the nearest categories from UK Labour Market statistics\footnote{\url{https://www.ons.gov.uk/employmentandlabourmarket/peopleinwork/employmentandemployeetypes/datasets/employeesandselfemployedbyindustryemp14}} as shown in Table~\ref{tab:female}.

\revision{We find that job ads in male-dominated job categories mention 3.20 soft skills on average, while ads in female-dominated job categories mention only 3.00 soft skills}. The difference in means is statistically significant (\revision{$p<0.001$; two-tailed $t$-test with unequal variances}). \revision{To predict the proportion of women in the category of a job ad, we used ordinary least squares (OLS) regression over job ads containing at least 3 different soft skills.}

\begin{table}[t!]
\centering
\caption{The percentage of women in job categories. Data from the UK Office for National Statistics (ONS), according to employment and labour market statistics (2018).}
\label{tab:female}
\begin{tabular}{l l l}
\toprule
\textbf{Job Category} & \textbf{ONS Category} & \textbf{\% of women} \\
\midrule
Social work Jobs & Human health \& social work activities & 80.62 \\
Healthcare \& Nursing Jobs & Human health \& social work activities & 80.62 \\
Charity \& Voluntary Jobs & Human health \& social work activities & 80.62 \\
Teaching Jobs & Education & 71.5 \\
\midrule
Property Jobs & Real estate activities & 57.6 \\
Legal Jobs & Public admin \& defence; social security & 56.02 \\
Creative \& Design Jobs & Other & 53.29 \\
Travel Jobs & Other & 53.29\\
Other/General Jobs & Other & 53.29 \\
Domestic help \& Cleaning Jobs & Accommodation and food services & 53.18 \\
Hospitality \& Catering Jobs & Accommodation and food services & 53.18 \\
Maintenance Jobs & Wholesale, retail \& repair of motor vehicles & 48.23 \\
Sales Jobs & Wholesale, retail \& repair of motor vehicles & 48.23 \\
Retail Jobs & Wholesale, retail \& repair of motor vehicles & 48.23 \\
Accounting \& Finance Jobs & Financial \& insurance activities & 46.45 \\
IT Jobs & Professional, scientific \& technical activities & 45.72 \\
Engineering Jobs & Professional, scientific \& technical activities & 45.72 \\
Scientific \& QA Jobs & Professional, scientific \& technical activities & 45.72 \\
HR \& Recruitment Jobs & Administrative \& support services & 44.4 \\
Customer Services Jobs & Administrative \& support services & 44.4 \\
Admin Jobs & Administrative \& support services & 44.4 \\
\midrule
PR, Advertising \& Marketing Jobs & Information \& communication & 29.77 \\
Consultancy Jobs & Information \& communication & 29.77 \\
Manufacturing Jobs & Manufacturing & 24.72 \\
Logistics \& Warehouse Jobs & Transport \& storage & 23.93 \\
Energy, Oil \& Gas Jobs & Mining, energy and water supply & 21.76\\
Trade \& Construction Jobs & Construction & 19.21 \\
\midrule
Graduate Jobs & - &N/A \\
Part time Jobs & - & N/A \\
\bottomrule
\end{tabular}
\end{table}

\revision{Table~\ref{tab:genderCoef} shows the soft skill clusters that are most predictive of female-dominated jobs (positive coefficients) and of male-dominated jobs (negative coefficients). Only those skill clusters that occurred more than 50 times and whose coefficient is statistically significant ($p < 0.01$) are shown. The table also indicates whether the reward associated with a soft skill is significant or not. The model obtained an $R^2$ score of 0.11.}

\begin{table}[t]
\centering
\caption{OLS regression results predicting the proportion of women using soft skill clusters as predictors. \revision{The first twelve soft skills are the strongest predictors for female oriented job ads (i.e. job ads for professions with a high proportion of women), while the last twelve rows correspond to the strongest predictors for male oriented ads (i.e. job ads for professions with a low proportion of women).
Many of the found predictors correspond to common gender stereotypes.
The third column lists the salary reward (\boldR, see Eq.~\ref{eq:r}), whilst the fourth shows the number of samples from the training set in which the skill clusters appear.\\}
}

\label{tab:genderCoef}
\begin{tabular}{l l l r}
\toprule
\textbf{Skill Cluster} & \textbf{Coefficient} & \boldR & \textbf{Count} \\
\midrule

ability to work with children & 0.192 & 0.3 & 370\\
delegation skills & 0.095&10.2\twoS &117\\
respectful & 0.090&-0.3 & 254\\
managerial skills & 0.083 & 3.0 &146 \\
reasoning skills & 0.072 &-10.6\twoS & 91\\
empathy & 0.067 & -1.3 & 576\\
ability to maintain confidentiality & 0.059 &-0.7 & 290\\
sensitivity & 0.056 & 3.0 &208\\
ability to adapt & 0.046 &4.2\twoS &635\\
dedication & 0.042 &-4.6\twoS &923\\
flexible with hours & 0.040 &-0.3 &1864\\
attentive & 0.039 &2.2 &124\\

\midrule

marketing skills & -0.046 &-0.7 &337\\
client skills & -0.045 &3.2\twoS &975\\
ability to win new business & -0.040 &2.2 &269\\
ability to lead project teams & -0.037 &3.7 &200\\
curious & -0.036 &4.1 &149\\
diligent & -0.035 &-0.8 &241\\
ability to present ideas & -0.034 &2.7 &149\\
courteous & -0.030 &0.6 &450\\
methodical & -0.028 &-1.1 &1035\\
attention to detail & -0.028 &-1.1 &7191\\
self starter & -0.026 &3.0\oneS &904\\
analytical skills & -0.026 &2.9\twoS &3972\\

\bottomrule
\addlinespace[1ex]
\multicolumn{3}{l}{\twoS$p<0.01$, \oneS$p<0.05$}
\end{tabular}
\end{table}

A high proportion of women in a job category is associated with soft skills such as \soft{empathy, respectful, sensitivity} and \soft{dedication}. Skills such as \soft{marketing skills, ability to win new business, ability to lead project teams} and \soft{analytical skills} are negatively associated with women's shares in job categories, meaning they predict soft skills mentioned more frequently in ads for male-dominated jobs. These results illustrate that with a few exceptions (e.g. \soft{delegation skills} and \soft{managerial skills}), \revision{the soft skills that are predictive of the job's gender composition are also closely associated to gender stereotypes.}

Thus, not only do skills associated with gender stereotypes about women potentially get lower rewards in labor markets (as suggested by Table~\ref{top_10_rewarding_skills}), but we further find that some soft skills, which are distinctive of the gender composition within a job, are also stereotyped as being female.
Put differently, not only does one \revision{potentially} get paid less if one is carrying out tasks connoted as being female, but occupations carried out mainly by women \revision{are also advertised making use of those skills that come about with wage penalties}. 

Our findings also suggest that there are two deviations from this pattern, i.e. \soft{delegation skills} and \soft{managerial skills}, which are soft skills that are associated with leadership (male) stereotypes but still predict a high proportion of women in an occupation. This finding, however, is in line with previous research, providing evidence that women will apply for leadership positions if the remaining part of the job ad is phrased using female stereotypes or gender neutral language~\cite{gaucher2011evidence, bem1973does, askehave2014gendered}.

\subsection{Occupational segregation and gender-stereotypical soft skills}
\label{sec:segregation}

To more systematically analyze the claim that the gender composition of an occupation is shaped by gender stereotypes, we mapped our soft skill clusters to a list of twenty personality characteristics desired in men and another twenty characteristics desired in women---the so-called Bem Sex Role Inventory~\cite{bem1974}. Out of these, we were able to map five feminine and seven masculine characteristics to similar soft skill clusters in our data, shown in Table~\ref{tab:stereotypes}.\footnote{Additionally, we found the following four matches: \soft{Act as a leader} $\rightarrow$ \soft{leadership}, \soft{Self-reliant} $\rightarrow$ \soft{confident}, \soft{Cheerful} $\rightarrow$ \soft{cheerful personality}, and \soft{Sympathetic} $\rightarrow$ \soft{sympathy}. These were, however, left out from our analysis since the former two soft skills had already been assigned to other similar stereotypes and the latter two have insufficient samples sizes of Count=3 and Count=4, respectively.} Based on the mappings, we set out to study the prevalence of the gender-stereotypical soft skills in job ads of female and male-dominated industries. The percentage of ads containing a skill within the ads from female- (male-) dominated industries is denoted by $P_f$ ($P_m$). In the last column of Table~\ref{tab:stereotypes} we show the percentage difference between these two percentages. A positive value means that the skill is used more in female-dominated industries and a negative value that it is used more in male-dominated industries.

\begin{table*}[t]
\centering
\caption{A mapping between stereotypes and soft skill clusters. 
The gender stereotypes listed by Bem~\cite{bem1974} that could be mapped to one of our soft skill clusters. On average, the feminine stereotypes are associated with a wage penalty ($r = -1.7$), whereas the masculine stereotypes are associated with a premium ($r = 2.6$). The percentage of job ads within female and male-dominated industries that mention a skill cluster are denoted by $P_f$ and $P_m$, respectively.} 

\label{tab:stereotypes}
\resizebox{\textwidth}{!}{
\begin{tabular}{c l l l c c c}
\toprule
& \textbf{Gender stereotype (Bem, 1974)} & \textbf{Mapped Skill Cluster} & $\boldR$ ($\%$) & $\mathbf{P}_f$ ($\%$) & $\mathbf{P}_m$ ($\%$) & $\frac{\mathbf{P}_f - \mathbf{P}_m}{\max(\mathbf{P}_f,  \mathbf{P}_m)} \times 100\%$ \\
\midrule
\parbox[t]{2mm}{\multirow{5}{*}{\rotatebox[origin=c]{90}{\textit{Feminine}}}}
& Compassionate & empathy & -1.3 & 0.94 & 0.12 & 87.1 \\
& Does not use harsh language & polite & -5.9\twoS  & 0.25 & 0.22 & 13.1 \\
& Loves children & ability to work with children & 0.3 &  2.13 & 0.07 & 96.8 \\
& Sensitive to the needs of others & sensitivity & 3.0 & 0.22 & 0.10 & 52.5 \\
& Warm & friendly personality & -4.6  & 0.11 & 0.07 & 38.4 \\
\midrule
& Average &  & -1.7 & 0.73 & 0.12 & 57.6 \\
\midrule
\parbox[t]{2mm}{\multirow{7}{*}{\rotatebox[origin=c]{90}{\textit{Masculine}}}}
& Ambitious & ambitious & 1.4 & 3.11 & 5.17 & -39.9 \\
& Analytical & analytical skills & 2.9\twoS  & 0.59 & 3.16 & -81.3 \\
& Assertive & confident & 0.5 & 6.39 & 6.09 & 4.7 \\
& Has leadership abilities & leadership & 7.4\twoS & 9.85 & 5.94 & 39.7 \\
& Independent & capability to work independently & 1.9 & 1.17 & 1.11 & 5.4 \\
& Makes decisions easily & make decisions & 3.0\twoS & 1.25 & 1.08 & 13.1 \\
& Self-sufficient & autonomy & 1.3 & 0.99 & 1.23 & -19.4 \\
\midrule
& Average &  & 2.6\twoS & 3.34 & 3.40 & -11.1 \\
\bottomrule
\addlinespace[1ex]
\multicolumn{3}{l}{\twoS$p<0.01$, \oneS$p<0.05$}
\end{tabular}
}

\end{table*}

All feminine skills are more prevalent in female-dominated industries, whereas for masculine skills the picture is not as clear. For instance, \soft{analytical skill} \revision{is} used more than five times more often in male-dominated industries, while \soft{leadership} is used almost \revision{twice as} often in female-dominated industries, although both of these skills are stereotypically masculine according to Bem~\cite{bem1974}. \revision{This finding, however, is in agreement with previous research, where evidence was found that although women will make inroads into occupations in which the skill set is in line with typically male features, this is not true the other way around~\cite{england2010gender,levanon2016persistence}. Hence, although women try to push into male-dominated occupations, men do not do the same with regard to female-dominated occupations}.

Our findings have implications for occupational sex segregation, that is, the unequal distribution of men and women across occupations in the labour market. Advertising female or male-dominated jobs in accordance with the associated gender stereotypes reproduces cultural beliefs about these stereotypes and upholds the gender-typicality of occupations. Previous research has shown that cultural beliefs about gender stereotypes influence self-assessment of men and women~\cite{correll2004,correll2001}. These biased self-assessments have been shown to be a crucial factor of career choices~\cite{correll2001}. \revision{Accordingly, empirical evidence employing experiments, suggest that if jobs are advertised using stereotypically male traits, women are less likely to think that they are suitable for the position~\cite{taris1998gender} and, hence, hesitate to apply. Thus, by illustrating that real jobs advertisements that include female stereotypes are dominated by women and vice-versa for men, we provide large-scale evidence that job ads can be seen as part of a leaky pipeline~\cite{shaw2018pipeline}, serving as the first sorting mechanism by which women are crowded out of male-dominated occupations at labor markets~\cite{askehave2014gendered,gaucher2011evidence, taris1998gender, born2010impact}}. 
 
The results thereby  
suggest the importance of gender stereotypes in the reproduction of occupational segregation, i.e. the demand-side, and the corresponding selection of men and women in different occupations. 
However, it is important to note that while our results establish a correlation between the usage of stereotypical soft skills and occupational segregation, studying the causal mechanisms between the two is beyond the scope of this paper. Nevertheless, this work supplements the much richer account of research examining the supply side of the unequal distribution of men and women across occupations, namely the influence of gendered individual preferences and respective assessments of one's own skills and capacities~\cite{busch2014supply,correll2001}, by 
showing a connection between the demand-side, i.e. job ads, and occupational segregation.

\subsection{Gendered Soft Skills and Salary}
\label{sec:divreward}

\revision{Results in the previous section illustrated that soft skills corresponding to gender stereotypes are associated with the gender composition of the job category. In what follows, we are going to examine to what extent these gendered soft skills are associated with wage premiums or penalties.}

Gender stereotypes may influence wages. More specifically, tasks that are linked to typically ``female'' responsibilities are often associated with wage penalties~\cite{england1994, kilbourne1994, england1992}. An explanation for the devaluation of ``female'' tasks is found in the ascribed lower status of women, i.e. gender status beliefs. Gender status beliefs are diffuse cultural beliefs on account of which men are rated more competent than women. These beliefs about women's lack in aptitude and competence are transferred to the labor market and thereby facilitate a devaluation of women and typically ``female'' tasks in the workplace~\cite{ridgeway1997}. \revision{Recent evidence, for instance, suggests that women are underrepresented in academic fields where practitioners believe that raw talent is needed in order to succeed. Women are simply seen as less brilliant than men and therefore not hired in academic segments where beliefs about the need for innate talent are salient~\cite{leslie2015expectations}}.

The rewards in Table~\ref{tab:genderCoef} illustrate that soft skills that correspond to gender stereotypes about women, such as \soft{respectful, empathy} and \soft{dedication} are predominantly associated with wage penalties (with the exception of \soft{sensitivity}). A similar pattern is found in Table~\ref{top_10_rewarding_skills}, where most of the soft skills related to stereotypes about women are associated with wage penalties, while the ones linked to leadership bring about wage premiums.
 
Hence, our study presents evidence on the devaluation of soft skills related to gender stereotypes based on a large-scale list of soft skills derived from real job ads and thereby confirms small-scale previous research, in which evidence was found that, net of individual labour-market-relevant characteristics such as work experience, \revision{single tasks tied to female gender stereotypes (such as \soft{nurturing}~\cite{england1994})} are associated with wage penalties~\cite{england1992, kilbourne1994}.

\revision{Regarding male-dominated jobs, our results show that soft skills that are associated with commonly shared stereotypes about men, such as \soft{analytical skill and self starter}~\cite{gaucher2011evidence}, predict statistically significant wage premiums.} Moreover, Table~\ref{tab:genderCoef} illustrates that leadership skills, which \revision{are} also stereotypically ascribed to men, do come with wage premiums (i.e., \soft{ability to win new business, ability to lead project teams}, and \soft{ability to present ideas)}. However, we find that leadership skills associated with female-dominated occupations such as \soft{delegation skills}, and \soft{managerial skills} are related to wage premiums \revision{as well}. This means that soft skills that are associated with a high share of women in an occupation are also more often related to wage penalties compared to soft skills that are associated with a high percentage of male incumbents. However, if soft skills \revision{required} in female-dominated occupations represent leadership skills they can also comprise wage premiums.

To \revision{further explore} the association between sex-typed gender stereotypes and wage penalties or premiums, we calculated the salary rewards $r$ of the soft skills clusters that we found congruent with the personality traits from the Sex Role inventory by Bem~\cite{bem1974}. The rewards are \revision{listed} in Table~\ref{tab:stereotypes}. We find that all masculine skills are associated with a positive reward, whereas 3/5 feminine skills are associated with a penalty. The average rewards for masculine and feminine skills are 2.6 and -1.7, respectively. This difference  is statistically significant (one-tailed $t$-test with equal variances; $p=0.014$). This suggests that stereotypically masculine character traits are valued more in the workplace than feminine character traits.

Based on the evidence provided we find that the devaluation of women is mainly realized via gender stereotypes, while skills associated with male stereotypes, i.e. leadership skills, do receive wage premiums. 

\section{Discussion and Conclusions}
\label{sec:conclusions}

This study examined soft skills in the labour market and showed that soft skills are a crucial component of job ads, especially of low-paid jobs and jobs in female-dominated professions and may therefore potentially perpetuate labour market inequalities.  
To explore how soft skills influence labor market outcomes, \revision{in particular wage premiums or penalties and gendered labour market composition}, we developed a semi-automatic approach for mining soft skills from job advertisements.

We would like to highlight three key findings of our study:
\begin{enumerate}
\item We found that not all soft skills are valued equally in the labour market, some are associated with wage premiums while others are linked to wage penalties.
\item Some soft skills are significant predictors of a job's gender composition. Utilizing solely soft skills, we can explain 11\% of the variation in the gender composition of job categories.
Soft skills that are associated with gender stereotypes, such as \soft{empathy} and \soft{sensitivity} for women, are significant predictors for a high percentage of women in the respective jobs, and vice versa is found for characteristics perceived as being ``male''.
\end{enumerate}
However, the selection of men and women into different occupations would in itself not be crucial for labour market inequality, as long as this segregation only implies that men and women work in different occupations and no other repercussions are attached. Previous research, however, has pointed out that wages paid in female-dominated occupations are lower than in male-dominated occupations~\cite{levanon2009occupational,mandel2013up,murphy2015feminization}. Sex segregation in labour market is thus perceived as being a crucial factor of perpetuating wage differentials between men and women. Therefore, our results suggest that gender stereotypical job ads serve as part of a leaky pipeline upholding gender wage inequality, \revision{by contributing} to a selection of women into lower paying occupations, \revision{on the basis of employing wording that discourages them to apply to higher paid male-dominated jobs in the first place}.

\par
\begin{enumerate}
\item[3] Typically ``female'' soft skills, i.e. prescribed stereotypes about women, are mostly associated with wage penalties, while soft skills associated with \soft{leadership}, and as such stereotypes that are associated with men, come with wage premiums---even after controlling for the job title and job category. 
\end{enumerate}

Although, by drawing on empirical research from psychology, we could explain which tasks are associated with being ``male'' or ``female'', we believe that certain soft skills, such as being \soft{respectful} and being \soft{curious} are probably important in any kind of job. Given this assumption, it is the more compelling to find that while the former is associated with a high percentage of women in an occupation and wage penalties, the latter comes about with wage premiums and is found in job ads for male-dominated occupations. This hints, as discussed, at a general devaluation of task carried out by women in labour markets.

\revision{One might wonder, if women could not simply apply for jobs that are advertised using ``male'' soft skills and thereby circumvent possible wage penalties. Current evidence however shows that the solution is not that simple: women are less likely to be successful when applying for a male-dominated job and when violating female gender stereotypes~\cite{davison2000sex, benard2010normative, rudman2001}.}

This study was not without limitations. Therefore next we discuss these restraints and briefly consider how these limitations can be addressed in the future research. 

\revision{First, distinguishing between when a given soft skill  is a necessity for a job or merely a useful asset is beyond the scope of this paper}. The accuracy of the soft skill detection \revision{method}, as well as the distinction of a soft skill being an asset or a necessity, could be improved by considering part-of-speech features.

Second, although we were able to account for a considerable degree of unobserved occupational heterogeneity by using matching techniques, in order to rigorously test the impact soft skills on wages, one would need to analyze if wage premiums or penalties associated with certain soft skills hold, net of individual labor-market-relevant attributes. \revision{More to the point: we believe that work experience and job tenure serve as relevant confounders in our study. The particularly large premiums for \soft{leadership} are very likely also connected to senior positions requiring professional expertise and longstanding on-the-job experience. While work experience is to some extent controlled by using the words of the job titles (e.g. \textit{senior} and \textit{intern}) as matching criteria, in some cases, the expected work experience can be indicated merely in the job description, which is not used for matching.} Given previous evidence that finds that tasks associated with being ``female'', such as ``nurturing skills'' do pose a penalty on wages, net of individual characteristics~\cite{england1994}, it is plausible that our results would be stable net of individual labor-market-relevant attributes as well. In future research this could be tested by linking the soft skills to individual survey data, \revision{which include measures of individual work experience}.

Regardless of these limitations, this study has made an important contribution to the impact of soft skills in the labour market. Combining computational methods as well as theoretical and empirical insights from economics, sociology and psychology enabled us to shed more light on how soft skills operate in the labour market. We showed that soft skills are a crucial component of job ads, especially of low-paid jobs and jobs in female-dominated professions. \revision{Furthermore, we found evidence} that soft skills \revision{are associated gender segregation across occupations} and \revision{reinforce} wage inequalities between men and women by rewarding typically ``male'' characteristics and penalizing ``female'' traits. 

Grugulis and Vincent~\cite[][p.599]{grugulis2009whose} put it this way: \enquote{When it is an individual character that is being judged, evaluations based on gender and race are far more likely}. Put differently, personal traits and characteristics, namely soft skills, are hard to evaluate and thus likely subjected to proxies such as gender or race and associated stereotypes, which in turn leads to discrimination. 
Our results support this observation, as they suggest that soft skill polarize labour market outcomes in terms of wages and occupational segregation. This polarization strikes women, as an already vulnerable group in labour markets, the hardest.

\section*{Acknowledgements}
We are grateful to Olaf Groh-Samberg, Karin Gottschall, \revision{Anne Busch-Heizmann}, Matti Nelimarkka, and \revision{two anonymous reviewers} for their invaluable feedback on previous versions of the article. \revision{All remaining errors are our own.}

\bibliographystyle{spbasic} 
\bibliography{new_bibtex}

\end{document}